# Phonon dephasing and spectral diffusion of quantum emitters in hexagonal Boron Nitride

Simon White[1], Connor Stewart[1], Alexander S. Solntsev[1], Chi Li[1], Milos Toth[1,2], Mehran Kianinia[1,2,*] and Igor Aharonovich[1,2]

1. School of Mathematical and Physical Sciences, Faculty of Science, University of Technology Sydney, Ultimo, New South Wales 2007, Australia
2. ARC Centre of Excellence for Transformative Meta-Optical Systems (TMOS), University of Technology Sydney, Ultimo, New South Wales 2007, Australia

Corresponding author Mehran.kianinia@uts.edu.au

***Abstract:*** *Quantum emitters in hexagonal boron nitride (hBN) are emerging as bright and robust sources of single photons for applications in quantum optics. In this work we present detailed studies on the limiting factors to achieve Fourier Transform limited spectral lines. Specifically, we study phonon dephasing and spectral diffusion of quantum emitters in hBN via resonant excitation spectroscopy at cryogenic temperatures. We show that the linewidths of hBN quantum emitters are phonon broadened, even at 5K, with typical values of the order of ~ one GHz. While spectral diffusion dominates at increasing pump powers, it can be minimized by working well below saturation excitation power. Our results are important for future utilization of quantum emitters in hBN for quantum interference experiments.*

Solid state quantum light sources are emerging as promising candidates for many applications in quantum technologies[1-3]. Among these sources, optically active point defects in hexagonal boron nitride are attracting considerable attention due to their extreme brightness, and high Debye Waller factor which means the majority of the photons are emitted into the zero phonon line (ZPL)[4-8]. While final defect assignments are still under debate[9], a number of recent experiments and theoretical papers hint at carbon related defects adjacent to a vacancy site in the hBN lattice[10,11]. In addition, numerous recent studies have shown that several defects in hBN exhibit spin dependent optical transitions, and exhibit optically detected magnetic resonance (ODMR), which is vital for their employment as solid state qubits and quantum sensors at the nano-scale [12-14].

For practical quantum photonic applications, where photon interference and generations of indistinguishable photons are required[15-18], it is important to characterize the coherent properties of the emitted photons. Specifically, studies of dephasing mechanisms[19,20], coherence and line broadening effects underpin the applicability of quantum emitters for photon interference experiments. Previous studies of hBN quantum emitters have revealed the emissions in hBN are broadly affected by spectral diffusion [21-27]. Preliminary resonant excitation experiments showed that observation of Fourier Transform limited lines is possible, but rather rare, as compared to other solid state emitters, such as diamond[28,29]. Some of the challenges stemmed from the fact that the level structure of the emitters is still poorly understood, and environmental effects in layered materials are strongly sample-dependent.

In this work, we employ coherent excitation spectroscopy at cryogenic temperatures to study the dephasing of quantum emitters in hBN. Importantly, our work focuses predominantly on coherent excitation (i.e. the excitation laser is on-resonance with the hBN emission). We find that even at low temperatures of 5K, the lines are predominantly broadened by phonon coupling. We also observe that spectral diffusion can be minimized by employing excitation powers well below the saturation power. We explain our results in the context of electron - phonon coupling and provide an important analysis for future experiments on two photon interference with quantum emitters in hBN.

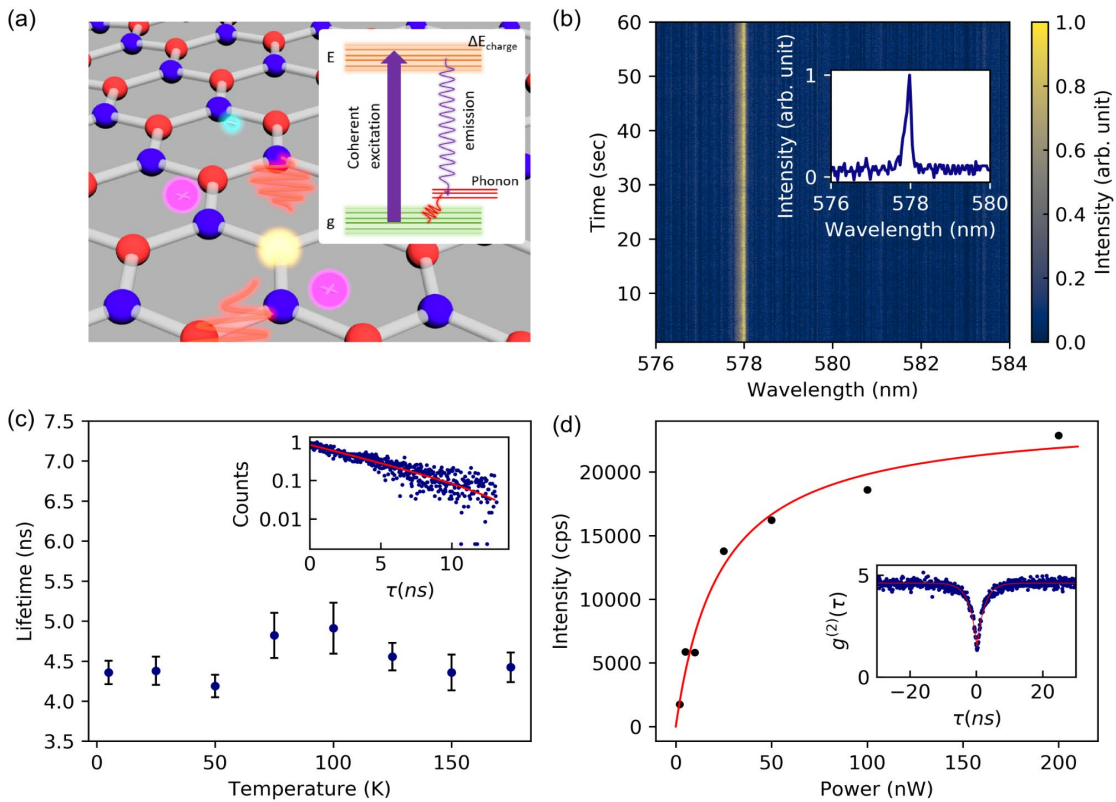

*Figure 1: Characterization of hBN single photon emitters. (a) Schematic representation of the hBN lattice with a point defect (yellow sphere). The emission from the defect is broadened by charge fluctuations and lattice phonons. The inset is a simplified electron energy diagram that illustrates dephasing pathways under resonant excitation. (b) PL spectra of hBN emitters under 532 nm excitation at 5 K. Each spectrum was recorded for 0.1s. The inset is a single PL spectrum of the same emitter. (c) hBN SPE lifetime showing no significant change vs temperature, under 512 nm pulsed excitation. Error bars are uncertainty of the fit (95% confidence interval). The inset demonstrates a representative lifetime measurement acquired*

*at 5 K. The red line is a single exponential fit of the data. (d) PL Saturation measurement of the hBN emitter under resonant excitation. The inset shows a resonant $g^2(\tau)$ measurement, recorded under 10 nW excitation power.*

The hBN flakes were mechanically exfoliated onto a silicon substrate from high-purity bulk hBN. The exfoliated hBN flakes were then treated with a $H_2$ plasma (10 minutes at 900 W, 100 sccm at 60 torr), after which the sample was annealed for 30 min in air at 850 ºC to remove any residual contaminants from the hBN surface. The sample was then placed under vacuum in a closed looped cryostat and cooled to 5 K.

The main dephasing mechanisms of the hBN emitters are shown in figure 1a. Coupling to lattice phonons (phonon dephasing) and random fluctuations of trapped charges in close proximity to the defect (spectral diffusion) are the main dephasing processes for all emitters in solid state hosts that result in broadening of the emission lines. To investigate a particular defect, we screen for a relatively bright emitter with spectrometer limited linewidth (<0.1 nm). Figure 1 (b) shows photoluminescence (PL) spectra of such an hBN emitter at 5 K under 1 mW excitation. The emitter does not exhibit any noticeable spectral diffusion as is illustrated by the time-series of PL spectra collected over a period of 60s. Next, we used a 512 nm pulsed laser with a repetition rate of 40 MHz to measure the lifetime of the emitter at temperatures in the range of 5 K to 150 K. Figure 1c shows that the emission decay rate is approximately constant over this temperature range. The total decay rate from the excited states, $\gamma = 1/\tau_0 = \gamma_r + \gamma_{nr}$, is, in general, a combination of temperature-independent radiative rate ($\gamma_r$), and a non-radiative rate ($\gamma_{nr}$) which depends on temperature. The dependence of the lifetime on temperature is typically described by the Mott–Seitz model for non-radiative relaxation, $\tau_0(T) = \tau_0(0) (1 + a \exp[-\Delta E / k_B T])^{-1}$, where $\Delta E$ is the activation energy, $a$ is the non-radiative relaxation strength parameter, and $k_B$ is the Boltzmann constant[30]. Our results suggest that non-radiative relaxation in hBN is not affected by phonons, at least within this temperature range (figure 1c). A lifetime of 4.4±0.1 ns was derived by fitting a single exponential to the data as is shown in the inset of Fig. 1c. This indicates a Fourier-Transform limited linewidth of ~ 36 MHz at 5 K from this emitter.

We next coherently excited the emitter by tuning a narrowband laser (linewidth < 100 kHz) to the emission energy of the emitter. For this measurement, the emissions into the phonon sideband was collected using a long-pass filter. First, we measured the saturation behavior of the emitter under resonant excitation as is shown in figure 2d. The solid line is a fit to the data with the equation ($I = I_{sat} p / (p + P_0)$), yielding a saturation power of $P_0 = 23.4$ nW and saturation intensity of $I_{sat} = 24.5$ kHz for this emitter. We recorded an autocorrelation curve from the emitter under 10 nW excitation power, as shown in the inset of figure 1d. note, we did not observe Rabi oscillations under excitation powers as high as 10 µW (i.e. well above saturation), indicating strong dephasing faster than the lifetime of the emission[22].

Next, we turn our attention to characterization of the dephasing processes of this emitter employing the resonant photoluminescence excitation (PLE) scheme. The PL intensity of the resonant excitation under a pump power of 7 nW is shown in Fig. 2 (a) fitted with Gaussian and Lorentzian functions (top part of the plot). The Lorentzian function fits better, and has substantially lower residuals compared to the Gaussian fit (figure 2a bottom panels). A Lorentzian shape indicates that the emission linewidth is homogeneously broadened, thus phonons are the dominant broadening mechanism even at ~ 5 K. The full width half maximum (FWHM) of the PLE spectrum in this case is 1.10 ± 0.04 GHz, which is significantly broader than the Fourier-Transform limited linewidth of 36 MHz estimated from the lifetime of the same emitter (see figure 1). Given this linewidth we would expect to see around 3% of photons to be emitted coherently such that they may be used for photon entanglement (Hong-Ou-Mandel) experiments. For a measured count rate below saturation around 10,000 cps we may expect around ~ 300 cps for identical photons. Note that the saturation measurements were recorded using phonon side band collection. Under cross polarization scheme[31], where ZPL photons can be collected a much higher count rate is expected, exceeding ~ 3000 counts/s for this particular system as an example.

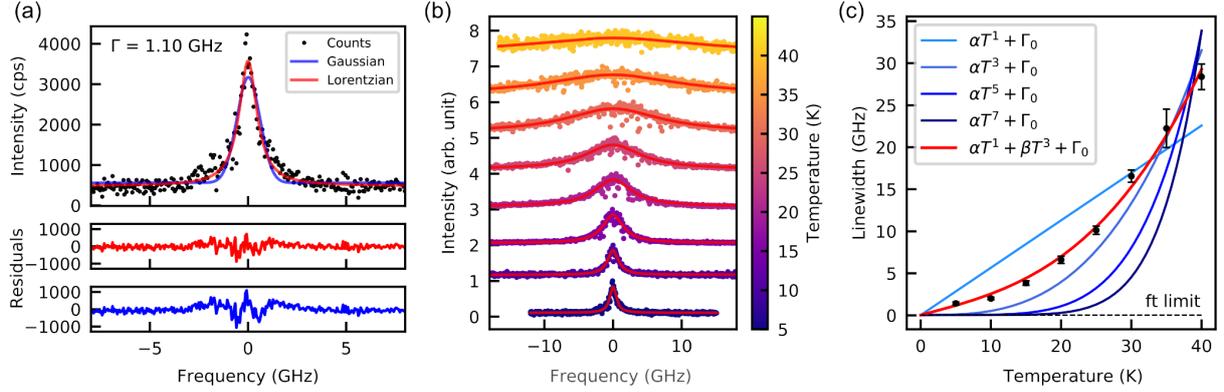

*Figure 2: Phonon-limited linewidth of hBN SPE. (a) Photoluminescence excitation (PLE) spectrum for hBN single emitter. The linewidth is fit with Gaussian and Lorentzian functions (top), with the corresponding residuals for each fit shown in the bottom panels. The Lorentzian fit is closest to the data and reveals a linewidth (FWHM) of 1.10+-0.04 GHz at 5 K. (b) Broadening of the emission linewidth is demonstrated as a function of temperatures due to increased interaction with phonons. Each spectrum is fitted with a Lorentzian function (solid red lines). (c) PLE linewidths (FWHM) as a function of temperature, with an error of 1 standard deviation from each Lorentzian fit. A model (solid red), which is a combination of $T^1$ and $T^3$ fits experimental data better than the other higher order polynomial fits.*

To explore the phonon-related PLE spectral broadening further, we show PLE spectra collected using a relatively low pump power of 7 nW over a temperature range of 4 to 40 K in Fig. 2 (b). The emission linewidth increases dramatically as the temperature rises, reaching nearly 30 GHz at 40 K, due to an increase in the interaction rate with phonons. We note that phonon broadening is still the dominant broadening mechanism over this temperature range, which we attribute to the relatively small corresponding change in thermal energy and hence a weak effect on spectral diffusion. As expected each spectrum shows a nearly perfect fit with a Lorentzian function, as shown in fig. 2(b).

In Fig. 2 (c), we plot the PLE FWHM linewidth against the sample temperature and fit the data with $aT^1$, $aT^3$, $aT^5$, $aT^7$ and $aT + bT^3$ curves [30,32,33]. As can be seen from the fittings, higher order polynomials do not fit as well as $aT + bT^3$. Closer to absolute zero, the dependence of the linewidth on the temperature is expected to be linear as a first-order approximation[34]. Since we are performing the measurements in the intermediate temperature range, a polynomial fit $aT+bT^3$ corresponds to the appropriate model. Above 20 K the linewidth scales as the cube of the temperature $\Gamma = (36 + 0.32\pm0.02 \cdot [T]^3)$ MHz. For low temperatures (<20 K), the behavior deviates from $T^3$ and is better approximated by a linear dependence on temperature $\Gamma = (36 + 220\pm29 \cdot [T])$ MHz. The observed $T$ and $T^3$ dependencies result from first- and second-order transitions due to electron–phonon interactions. The fits with higher-order polynomials $T^5$ and $T^7$ do not match our data well which indicates that in our system degenerate electronic states are not dominant, Jahn-Teller effect is small, the effect of strain is low, and the inelastic Raman process is not prominent[30,35,36]. This phonon broadening also indicates that further cooling (below 4 K) would enable further narrowing and may enable Fourier-Transform limited linewidth for such SPEs in hBN.

Next, we explore the dependence of the PLE emission as a function of pump power at 5 K to evaluate spectral diffusion of the emitter. Fig. 3 (a) shows a series of 8 individual scans at pump power of 100 nW, that results in a significant spectral diffusion. At the bottom of fig. 3(a), the integrated spectrum is shown and has a broadened linewidth of ~ 10 GHz. On the other hand, figure 3 (b) shows that when the power is reduced to 7 nW the spectral diffusion is significantly reduced to sub-GHz level.

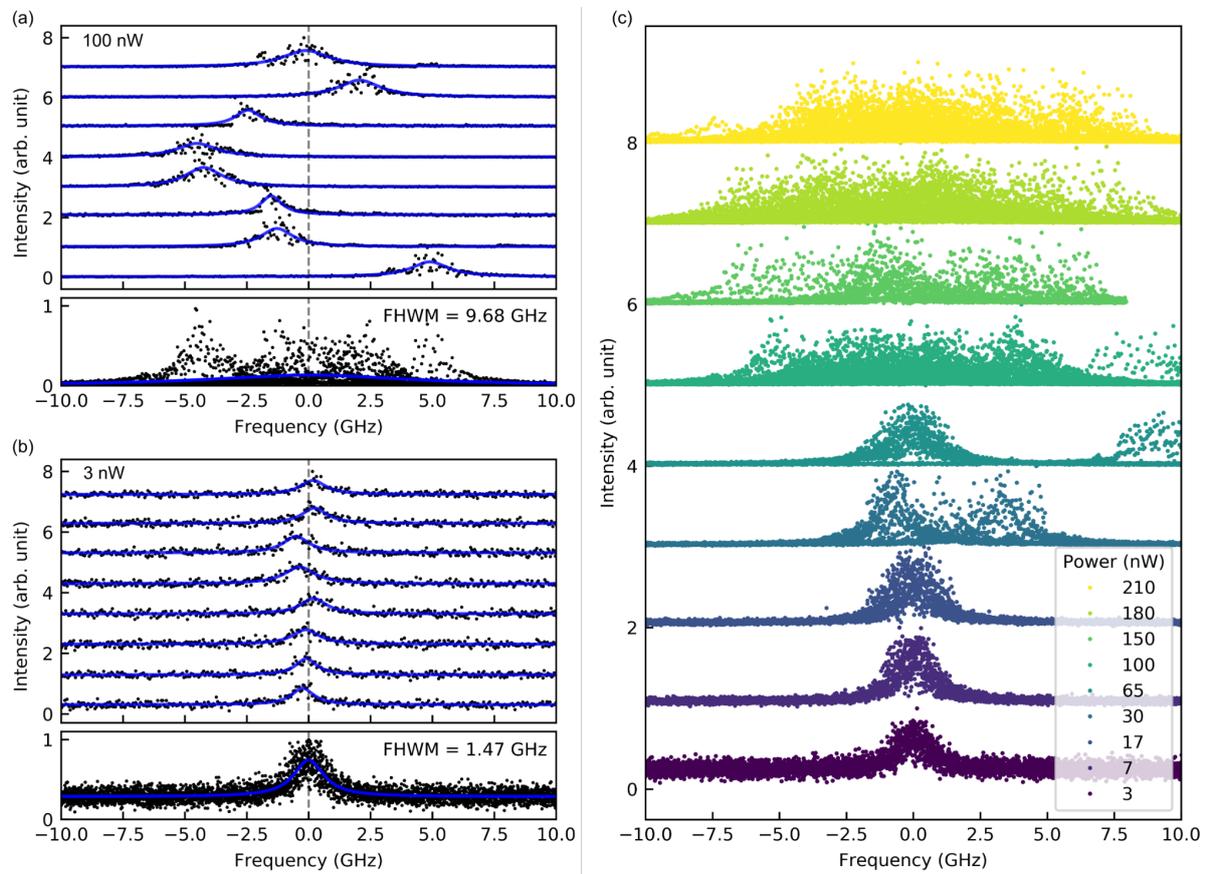

*Figure 3: Power-dependent spectral diffusion characterized by PLE spectroscopy. (a) Individual PLE scans at a pump power of 100 nW (top) and the corresponding integrated PLE spectrum (bottom) showing significant spectral diffusion. (b) Individual PLE scans at a pump power of 7 nW (top) and the corresponding integrated PLE spectrum (bottom) showing negligible diffusion (c) Integrated PLE spectra at pump powers in the range of 3 nW to 210 nW, showing that the integrated linewidth increasing with pump power.*

This measurement is consistent with the temperature dependent measurements which show that the linewidth, at a pump power of 7 nW, is broadened due primarily to interactions with phonons and to a lesser extent by spectral diffusion. Interestingly, we also note that the linewidth for individual scans maintains a similar linewidth to the low power scans (~ 1 GHz) signifying the dominant broadening mechanism of the line is still homogeneous. This also opens the possibility that active field modulation, post selection technique or charge depletion may play a role in enabling bright coherent emission from such hBN source[37,38]. By efficient decoupling of the emitter from its local environment one could avoid spectral diffusion and enable high power excitation and emission[38,39].

To reveal the full spectral diffusion behavior of the emitter, we recorded number of PLE scans using excitation powers from 3 to 210 nW, as shown in fig. 3 (c). Spectral diffusion is seen to significantly increase at a transition power around 30 - 65 nW, which corresponds with the saturation power, and stable coherent excitation becomes difficult. These results demonstrate that to use such hBN emitters for quantum interference and enable viable levels of indistinguishability between consecutive photons, it is essential to use excitation powers below saturation and/or lower temperatures.

To summarize, in this study we characterize the significant dephasing and spectral broadening mechanisms in an hBN single photon emitter under resonant excitation. We find that the resonant linewidth, even at cryogenic temperatures, is dominated by phonon broadening and results in linewidths of ~ 1 GHz. We also see that degenerate electronic states and strain do not play a significant role in phonon dephasing. We further showed that spectral diffusion can be minimized by employing excitation powers well below saturation. Overall, the brightness of the emitters exemplify that emission rate in excess of 200 MHz with bandwidth of less than 1 GHz.

The work opens exciting opportunities for quantum interference experiments with defects in hBN. Development of cross polarization schemes to collect ZPL photons should be implemented using waveguide structures[40-43]. This is certainly within reach with the currently available nanofabrication techniques and will enable substantially more photons. Our results also infer that extended cooling below 4 K can enable further narrowing of the spectral linewidth and may enable an approach to generate indistinguishable photons on demand. Finally, established tuning techniques[44,45] can be utilized to not only stabilize spectral diffusion, but tune two distinct hBN emitters into same resonance, thus paving way to generation of remote indistinguishable photons.

**Appendix**

1- Sample preparation:

The hBN used in this work was exfoliated from high crystal bulk hBN and then transferred to $SiO_2$/Si substrate with PDMS. A pre-annealing of 500C is conducted to remove the transfer residuals. The sample is then etched for 10 min in 900W hydrogen plasma at 60 torr using a microwave plasma deposition system(SEKI AX5100) to create defects on the flake surface. Finally, the sample was annealed for 30 min in a tube furnace (Lindberg/Blue M™) in air.

2- Optical measurement:

The sample was mounted on the cooling stage of an Attocube Attodry800 closed system cryostat, placed under vacuum, and cooled to 4.5 K. Optical measurements were then performed using a custom scanning confocal microscope with a 0.82 NA vacuum compatible objective mounted inside the cryostat (also at 4.5 K). A tunable dye laser (Sirah Matisse 2 DS) with linewidth around 100 kHz was used for resonant excitation and a 532nm diode laser (Laser Quantum GEM) was used for off-resonant excitation. For photoluminescence excitation measurements a scan rate of 1 GHz/s was used to tune dye laser wavelength over the ZPL, then the phonon sideband emission from SPE was filtered using a 580 nm long pass filter, coupled via a single mode fiber, and detected using an Avalanche Photodiode (Excelitas SPCM-AQRH). Second order correlation measurements were performed using a fiber beam-splitter and a TimeTagger20, and lifetime measurements were taken using 40 MHz pulsed 512 nm laser (PiL051X™, Advanced Laser Diode Systems GmbH) and TimeTagger20.

**Acknowledgement**
We acknowledge the Australian Research Council (CE200100010, DP190101058) and the Asian Office of Aerospace Research and Development (FA2386-20-1-4014) for the financial support.

**Conflict of Interest**
The authors declare no conflicts of interest